\documentclass[11pt,twoside,onecolumn]{article}
\usepackage[]{latexsym,amsmath,amssymb,graphicx,epsfig}
\newcommand{\be}{\begin{eqnarray}}
\newcommand{\ee}{\end{eqnarray}}
\newcommand{\ud}{\mathrm{d}}

\newcommand{\lp}{\ell_{\rm p}}
\newcommand{\mpl}{M_{\rm p}}
\newcommand{\mh}{M_{\rm H}}
\newcommand{\md}{M_{(5)}}

\newcommand{\Mpl}{M_{\rm p}}
\newcommand{\Meff}{M_{\rm eff}}
\newcommand{\Mc}{M_{\rm c}}
\newcommand{\Mh}{M_{\rm H}}
\newcommand{\ld}{\ell_{(5)}}
\newcommand{\mew}{M_{\rm ew}}

\newcommand{\mc}{M_{\rm c}}
\newcommand{\rc}{r_{\rm c}}

\newcommand{\rh}{R_{\rm H}}

\title{\bf Exploring the bulk of tidal charged micro-black holes}
\author{Roberto~Casadio$^a$\thanks{E-mail: casadio@bo.infn.it}$\ $
and
Octavian Micu$^b$\thanks{E-mail: octavian.micu@tu-dortmund.de}
\\
\\
{\em $^a$Dipartimento di Fisica, Universit\`a di
Bologna and I.N.F.N., Sezione di Bologna,}
\\
{\em via Irnerio~46, I-40126 Bologna, Italy}
\\
\\
{\em $^b$Fakult\"at f\"ur Physik, Technische Universit\"at
Dortmund,}
\\
{\em D-44221 Dortmund, Germany}
\\
\\
{ Preprint number: DO-TH-10/02}}
\date{}
\begin{document}
\maketitle
\begin{abstract}
We study the bulk corresponding to tidal charged brane-world black holes.
We employ a propagating algorithm which makes use of the
three-dimensional multipole expansion and analytically yields the metric elements
as functions of the five-dimensional coordinates and of the ADM mass,
tidal charge and brane tension.
Since the projected brane equations cannot determine how the charge
depends on the mass, our main purpose is to select the combinations
of these parameters for which black holes of microscopic size possess
a regular bulk.
Our results could in particular be relevant for a better understanding
of TeV-scale black holes.
\par
\null
\par
\textit{PACS - 04.50.Gh, 04.25.dg}
\end{abstract}
\section{Introduction}
\setcounter{equation}{0}
Several models with extra spatial dimensions~\cite{arkani,RS}
are nowadays available, in which Standard Model fields are
confined to a four-dimensional (thin) hypersurface
(the brane) embedded in the higher-dimensional space-time
(the bulk).
The existence of extra dimensions and a sufficiently small
fundamental scale of gravity would then allow for the possibility
to produce microscopic black holes~\cite{bhlhc,CH,BHreview,gingrich}
at the Large Hadron Collider (LHC).
\par
In this paper we shall, in particular, consider the
Randall-Sundrum (RS) brane-world of Ref.~\cite{RS}.
Our world is thus a three-brane
(with coordinates $x^i$, $i,j=0,\ldots,3$)
embedded in a five-dimensional bulk with the metric
\be
\ud s^2
=
e^{-\sigma\,|z|}\,g_{ij}\,\ud x^i\,\ud x^j
+ \ud z^2
\ ,
\label{g5}
\ee
where $z$ runs along the fifth dimension and
$\sigma^{-1}$ is a length determined by the brane tension.
This parameter relates the four-dimensional Planck mass
$\mpl$ to the five-dimensional gravitational mass $\md$
and one can have $\md\simeq 1\,$TeV$/c^2$
(for bounds on $\sigma$, see, e.g., Ref.~\cite{harko})
and black holes with mass in the TeV~range.
Note that, experimental limits require $\md\gtrsim 1\,$TeV,
but there is no strong theoretical evidence that places
$\md$ at any specific value below $\mpl$.
The brane must also have a thickness, which we denote by
$L$, below which deviations from the four-dimensional
Newton law occur.
Current precision experiments require that
$L \lesssim 44\,\mu$m~\cite{Lbounds},
whereas theoretical reasons imply that
$L\gtrsim\ld\simeq \lp\,\mpl/\md\simeq 2\cdot 10^{-19}\,$m.
In this context, compact sources on the brane,
such as stars and black holes, have been investigated
extensively.
However, their description has proven rather complicated and
there is little hope to obtain analytic solutions such as
those found with one dimension less~\cite{emparan}.
The present literature does in fact provide solutions on
the brane~\cite{maartens,germani,cfm}, perturbative results
over the RS background~\cite{katz,tanaka} and numerical
treatments~\cite{shiro}.
(For recent reviews, see Refs.~\cite{BHreview,bwbh}.)
\par
In this paper we investigate the bulk for the specific family of
asymptotically flat, static and spherically symmetric solutions on
the brane found in Ref.~\cite{maartens}.
These metrics, besides the Adler-Deser-Misner (ADM) mass
$M$, depend on an apparently free extra parameter,
the so called tidal charge $q$.
Our main aim is thus to restrict $q$ by requiring the bulk be ``regular'',
meaning that it can only contain the compact extension of the brane
horizon and no signs of other real singularities corresponding to physical
sources extending off-brane.
The method we shall use was introduced in Ref.~\cite{cm}
and will be briefly reviewed in the next Section, along with
the results previously obtained for candidate black holes
of astrophysical size.
In Section~\ref{tcbh}, we shall then apply the method to
the case of microscopic black holes.
Regular bulks will then be obtained only for a tidal charge
larger than the ADM mass but smaller than the (inverse of the)
brane tension.
Incidentally, this picture is compatible with the phenomenologically
allowed cases, as per the analysis of Ref.~\cite{cfhm}.
\par
We set the brane cosmological constant to zero by fine
tuning the bulk cosmological constant $\Lambda$ to the brane
tension $\sigma$, i.e.~$\Lambda=-4\,\sigma^2$~\cite{RS,shiromizu}
and use units with $1=c=\hbar=\mpl\,\lp=\ld\,\md$,
where $\mpl\simeq 2.2\cdot 10^{-8}\,$kg and
$\lp\simeq 1.6\cdot 10^{-35}\,$m
are the Planck mass and length related to the
four-dimensional Newton constant $G_{\rm N}=\lp/\mpl$.
In our analysis we shall consider only the five-dimensional RS~scenario
with $\md\simeq \mew\simeq 1\,$TeV ($\simeq 1.8\cdot 10^{-24}\,$kg),
the electro-weak scale, corresponding to the length
$\ld\simeq 2.0\cdot 10^{-19}$m.
\section{Reconstructing the bulk}
\setcounter{equation}{0}
We start by reviewing the algorithm introduced in Ref.~\cite{cm}.
On projecting the five-dimensional vacuum Einstein equations
$^{(5)}R_{\mu\nu}=\Lambda\,g_{\mu\nu}=-4\,\sigma^2\,g_{\mu\nu}$
(with $\mu,\nu=0,\ldots,4$)
onto the brane and introducing Gaussian normal
coordinates $x^i$ and $z$ ($z=0$ on the brane),
one obtains the constraints
\be
\left.^{(5)}R_{iz}\right|_{z=0}=
\left.^{(4)}R\right|_{z=0}=0
\ ,
\label{Deq}
\ee
where $^{(4)}R$ is the four-dimensional Ricci scalar
and use has been made of the necessary junction
equations~\cite{israel}.
Eqs.~\eqref{Deq} are analogous to the momentum and
Hamiltonian constraints in the ADM decomposition and
select admissible field configurations along a given
hypersurface of constant $z$.
Acceptable configurations are then ``propagated''
off-brane by the remaining Einstein equations,
\be
^{(5)}R_{ij}
=-4\,\sigma^2\,g_{ij}
\ .
\label{Ein2}
\ee
The above ``Hamiltonian'' constraint is weaker
than the purely four-dimensional vacuum equations
$R_{ij}=0$, being equivalent to $R_{ij}=E_{ij}$,
where $E_{ij}$ is (proportional to) the (traceless)
projection of the five-dimensional Weyl tensor on
the brane~\cite{shiromizu}.
We then consider five-dimensional metrics of the
form~\eqref{g5} with
\be
ds^2_{(4)}
\equiv
e^{-\sigma\,|z|}\,g_{ij}\,\ud x^i\,\ud x^j
=
-N(r,z)\,\ud t^2+A(r,z)\,\ud r^2+R^2(r,z)\,\ud\Omega^2
\ ,
\label{g}
\ee
where $\ud\Omega^2\equiv\ud\theta^2+\sin^2\theta\,\ud\phi^2$
and $N$, $A$ and $R$ are functions to be determined.
The momentum constraint is then identically solved
and the ``Hamiltonian'' constraint reads
\be
2\left(\frac{N_{\rm B}'}{N_{\rm B}}\right)'
+\left(\frac{N_{\rm B}'}{N_{\rm B}}+\frac{4}{r}\right)
\left(\frac{N_{\rm B}'}{N_{\rm B}}
-\frac{A_{\rm B}'}{A_{\rm B}}\right)
=\frac{4}{r^2}\left(A_{\rm B}-1\right)
\ ,
\label{H}
\ee
where the subscript B means that all functions are
evaluated on the brane at $z=0$,
$\ '\equiv \partial/\partial r$
and we set $R_{\rm B}=r$ thanks to (four-dimensional)
spherical symmetry~\cite{wald}.
\subsection{Propagating algorithm}
The bulk metric will be determined in three steps:
\begin{enumerate}
\item
\label{step1}
choose a metric of the form~\eqref{g} whose projection
on the brane,
\be
\left.\ud s^2_{(4)}\right|_{z=0}
=-N_{\rm B}(r)\,\ud t^2+A_{\rm B}(r)\,\ud r^2+r^2\,\ud\Omega^2
\ ,
\ee
solves the constraint~(\ref{H});
\item
\label{step2}
expand this metric in powers of $1/r$ (four-dimensional
{\em multipole expansion}) to order $n$,
\be
\left.
\begin{array}{c}
N_n(r,z) \\
A_n(r,z) \\
R^2_n(r,z)
\end{array}\right\}
\equiv\sum_{k=0}^n\,\frac{1}{r^k}
\,
\left\{
\begin{array}{l}
n_k(z) \\
a_k(z) \\
r^2\,c_k(z)
\end{array}\right.
\ ,
\label{series}
\ee
where $n_k(0)$, $a_k(0)$ and $c_k(0)$ reproduce the
solution chosen at step~\ref{step1} (to order $n$):
\be
\sum_{k=0}^n\,\frac{1}{r^k}
\,
\left\{
\begin{array}{l}
n_k(0) \\
a_k(0) \\
r^2\,c_k(0)
\end{array}\right\}
=
\left\{
\begin{array}{c}
N_{\rm B}(r) \\
A_{\rm B}(r) \\
r^2
\end{array}\right\}
+{\mathcal O}\left(\frac{1}{r^{n+1}}\right)
\ ;
\label{bou}
\ee
\item
\label{step3}
substitute the sum~\eqref{series} into Eq.~\eqref{Ein2}
and integrate analytically in $z$ the (three) equations
thus obtained for the functions $n_n(z)$, $a_n(z)$ and
$c_n(z)$.
\end{enumerate}
This procedure turns out to be particularly convenient
because it converts the Einstein equations~\eqref{Ein2}
into second order ordinary differential equations of
the form
\be
\frac{\ud^2 f_n}{\ud z^2}-\sigma^2\,f_n=F_{k<n}
\ ,
\label{f_eq}
\ee
where $f_n$ is any of the functions $n_n(z)$, $a_n(z)$ and
$c_n(z)$ ($n\ge 1$), and $F_{k<n}(z)$ a functional
of the lower order terms $f_{k<n}$'s and their first and
second derivatives.
The relevant boundary conditions are given by
Eq.~\eqref{bou} for $f_n(0)$ and the
junction conditions~\cite{israel}
\be
\left.\frac{\ud f_n}{\ud z}\right|_{z=0}=-\sigma\,f_n(0)
\ .
\ee
For $n=0$ one has a system of three coupled second
order ordinary differential equations for the $f_0$'s
and the corresponding Cauchy problem is uniquely
solved by the usual warp factor, $f_0=\exp(-\sigma\,z)$.
The $F_{k<n}$'s then turn out to be such that
the Cauchy problem at order $n$ admits analytical solutions
and one can determine the functions $f_n$ recursively.
\par
Except for the algebraic constraints following from
Eq.~\eqref{H}, the coefficients $n_k(0)$ and $a_k(0)$,
which are related to the shape of the source,
can be chosen at will.
However, it is in general difficult
to pinpoint one parameter (among the coefficients of the
multipole expansion) whose ``smallness'' guarantees
that orders higher than $n$ be negligible.
Ideally, the resulting bulk metric is reliable
for those values of $r$ and $z$ such that
\be
\frac{|f_{n+1}(z)|}{r^{n+1}}\ll
\left|\sum_{k=0}^{n}\,\frac{f_k(z)}{r^k}\right|
\ ,
\label{conv}
\ee
for given values of the parameters $n_k(0)$ and $a_k(0)$.
In general, for a given $z$, such a condition will only be
satisfied for sufficiently large $r$.
This implies that different choices
of $n_k(0)$ and $a_k(0)$ might lead to bulk space-times
which remain indistinguishable because the differences
are confined within too small a region close to $r=z=0$.
Conversely, by looking at the bulk metric in our approach,
we may not be able to distinguish the correct brane source
from other sources, e.g., those which spread slightly
into the bulk.
\subsection{Astrophysical examples}
As examples of brane metrics, Ref.~\cite{cm} considered
the solutions given in Refs.~\cite{maartens,germani,cfm}
which can be expressed in terms of the ADM mass
$M=r_{\rm h}/2$ ($r_{\rm h}$ is the event horizon)
and the post-Newtonian parameters
$\beta=\gamma=1+{1\over 3}\,\eta$~\cite{will}
on the brane.
Exact Schwarzschild on the brane, $\eta=0$,
is called black string (BS)~\cite{chamblin} and suffers
of serious stability problems~\cite{chamblin,gregory}.
One can therefore argue that brane-world black holes
have $\eta\not=0$ and cases with $\eta<0$
are favored, since $\eta>0$ implies anti-gravity effects~\cite{cfm}.
For astrophysical sources of solar mass size, Ref.~\cite{cm}
made use of the typical values
$M=10^7\,\sigma^{-1}\sim 1\,{\rm km}$ and
$\eta=-10^{-4}$.
In such a range,
\be
M\,\sigma\gg 1
\ ,
\label{astro}
\ee
(with $|\eta|\ll 1$),
one finds a qualitatively identical behavior for all
brane metrics in Refs.~\cite{maartens,germani,cfm}.
In particular, positive exponentials appear in the
metric functions, which are non-perturbative in
$z$, and make the expansion in $1/r$ preferable
(or complementary) to the expansion for small $z$.
\par
\begin{figure}[t!]
\centering
\epsfxsize=2.9in
\epsfbox{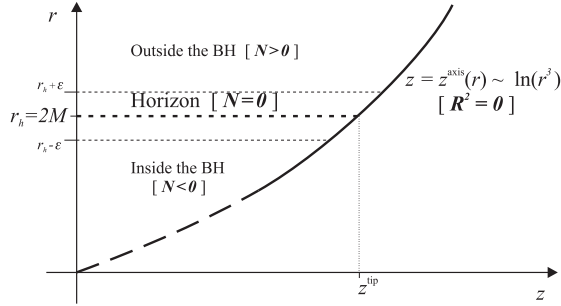}
\caption{Qualitative picture of bulk structure.
\label{mazza}}
\end{figure}

For $\eta<0$, it was found that, for
$n\ge 3$ and $r>0$, there exists a
$z=z_n^{\rm axis}(r)$ such that
$R_n^2(r,z_n^{\rm axis})=0$.
Since $4\,\pi\,R^2$ is the proper area of the sphere
$t=r=z=\,$constant, this suggests that the axis of
cylindrical symmetry is given by a line
$z=z^{\rm axis}(r)$.
Although the condition~\eqref{conv}
fails for $z=z_n^{\rm axis}$,
in the physically interesting range, the $1/r$ expansion
yields rather stable values of $z_n^{\rm axis}$
in a wide span of $n$, stability improves for larger
values of $r$ and becomes very satisfactory for
$r\gtrsim r_{\rm h}$.
For example, for $r\gg r_{\rm h}$, one finds
\be
z_5^{\rm axis}
\simeq
\frac{1}{2\,\sigma}\,
\ln\left(\frac{3\,\sigma^2\,r^3}{-\eta\,M}\right)
\ ,
\label{zaxis}
\ee
which numerically agrees fairly well with
$z_{5<n\le 19}^{\rm axis}$.
If the horizon closes in the bulk, then
it must cross the axis of cylindrical symmetry at
a point (the ``tip'') of finite coordinates
$(r^{\rm tip},z^{\rm tip})$, which can be obtained
approximately by solving
\be
N_n(r_n^{\rm tip},z_n^{\rm tip})
=
R^2_n(r_n^{\rm tip},z_n^{\rm tip})
=
0
\ .
\label{tip}
\ee
For $n=5$, one finds
\be
z_5^{\rm tip}
\simeq
\frac{1}{\sigma}\,
\ln\left(\frac{M\,\sigma}{\sqrt{-\eta}}\right)
\ .
\label{ztip}
\ee
For large values of $n$, one can only solve Eqs.~\eqref{tip}
numerically and finds a good parameterization for the horizon
is given by $r\simeq r_{\rm h}$ and
$0\le z\lesssim z^{\rm axis}(r_{\rm h})\simeq z^{\rm tip}$
(see Fig.~\ref{mazza}), which strongly suggests that the horizon closes in the
bulk, in accord with numerical analysis~\cite{shiro}.
One can also estimate how flattened the horizon
is towards the brane by comparing the proper length of a
circle on the brane-horizon, ${\mathcal C}_{||}=2\,\pi\,r_{\rm h}$,
with the length of an analogous curve perpendicular to
the brane, ${\mathcal C}_{\perp}\simeq 4\,z^{\rm tip}$.
Since their ratio is huge, one can in fact speak of a
``pancake'' horizon as was suggested, e.g., in Ref.~\cite{katz}.
In particular, the area of the (bulk) horizon is approximately equal
to the four-dimensional (brane) expression~\footnote{The
fundamental (possibly TeV scale) five-dimensional gravitational
coupling $G_{(5)}\sim G_N/\sigma$, where $G_N$ is the
four-dimensional Newton constant \cite{RS}.
Thus, from~\eqref{area}, one has
$^{(5)}{\mathcal A}/G_{(5)}
\sim
M^2/G_N
\sim\,^{(4)}{\mathcal A}/G_N$.},
\be
^{(5)}{\mathcal A}
\simeq
4\,\pi\,\int_0^{z_5^{\rm tip}} R^2(r_{\rm h},z)\,\ud z
\simeq
\frac{16\,\pi}{\sigma}\,(2\,M)^2
\ ,
\label{area}
\ee
where we again used $M\,\sigma\gg 1$.
Drawing upon the above picture, in particular the
crossing of lines of constant $r$ with the axis of
cylindrical symmetry at finite $z$, one can infer
that the Kretschmann scalar in these space-times
is well-behaved~\cite{cm}, contrary to
the BS~\cite{chamblin}.
\par
We conclude by mentioning that cases with $\eta>0$
show a very different qualitative behavior.
One finds that $R_n^2(r,z)$ is generically a
(monotonically) increasing function of $z$ for all
(sufficiently large) values of $r$, as one would
indeed expect on a negative tension
brane~\cite{shiromizu}.
However, for any $r_1$, $r_2>0$ there now exists
$z^*_n(r_1,r_2)$ such that
$R^2_n(r_1,z^*_n)=R^2_n(r_2,z^*_n)$, i.e.,
space-like geodesics of constant $r$ display
caustics and the Gaussian coordinates $(r,z)$
do not cover the whole bulk~\cite{wald}.
\section{Tidal charged black holes}
\label{tcbh}
\setcounter{equation}{0}
We now proceed to analyze the brane metric found in Ref.~\cite{maartens},
\be
\left.\ud s^2_{(4)}\right|_{z=0} =
- N_{\rm B}\,\ud t^2 + N_{\rm B}^{-1}\,\ud r^2 + r^2\,\ud\Omega^2
\ ,
\label{tidal}
\ee
with
\be
N_{\rm B}=1-\frac{2\,M}{r}-\frac{Q}{r^2}
\ ,
\ee
and~\cite{cfm}
\be
\eta=-\frac{2\,Q}{M^2}
\ .
\ee
The tidal charge $Q$ is assumed positive, so that there is one horizon at
\be
\rh=\left(M+\sqrt{M^2+Q}\right)
\ .
\label{rhq}
\ee
Moreover, the Einstein equations projected onto the brane do not relate the
ADM mass to the tidal charge, and $Q$ therefore appears as a free parameter.
We however expect that $Q$ vanishes when $M$ does, and the former should thus
be a function of the latter~\cite{CH}.
Nonetheless, we shall apply our propagating algorithm to the metric~\eqref{tidal}
with independent $M$ and $Q$.
\subsection{TeV-scale black holes}
Like in Refs.~\cite{CH,cfhm,cfh}, we are here interested in
microscopic black holes with a mass $M$ close to the assumed
value of $\md$.
Since $\sigma^{-1}\gtrsim\lp$, for such small black holes
the opposite of the condition~\eqref{astro} holds, namely
\be
M\, \sigma\ll 1
\ .
\label{csigmanob}
\ee
\par
The above relation allows one to constrain the relative
strength of tidal effects with respect to the Newtonian potential.
This kind of analysis was indeed performed earlier in Ref.~\cite{cfhm},
and we here review it briefly.
First of all, we recall that $Q$ is related to the charge $q$ of
Refs.~\cite{CH,cfhm,cfh} by
\be
Q=q\,\lp^2\left(\frac{\mpl}{\md}\right)^2
=\lp^{2-\beta}\left(\frac{\mpl}{\md}\right)^{\alpha+\beta+2}M^\beta
\ ,
\label{Q}
\ee
where $\alpha$ and $\beta>0$ are parameters.
We then note that the tidal term in the metric dominates over the usual General
Relativistic term for $r\lesssim \rc$, with
\be
\rc\simeq
\lp^{~2-\beta}\,\left(\frac{\Mpl}{\md}\right)^{\alpha+\beta+2}\,M^{\beta-1}
\ .
\label{rc}
\ee
This implies that $\rc$ must be shorter than the length scale $L$
above which corrections to the Newtonian potential have not yet been detected.
That is, we impose
\be
\rc\ll L
\ ,
\label{rcc}
\ee
and the black hole is therefore ``small'' provided
\be
\rh\ll\rc\ll L
\ .
\label{sbh}
\ee
In fact, for $\rh\ll\rc$, the horizon radius can be approximated by the tidal part
of Eq.~\eqref{rhq},
\be
\rh\simeq
\lp^{~\frac{2-\beta}{2}}\,\left(\frac{\Mpl}{\md}\right)^{\frac{\alpha+\beta+2}{2}}
M^{\frac{\beta}{2}}
\ ,
\label{smallRh}
\ee
otherwise, it approaches the usual four-dimensional expression
$\rh\simeq 2\,M$.
The effective four-dimensional Euclidean action~\cite{CH,gergely},
for small black holes, can be approximated by
\be
S_{(4)}^{\rm E}
=
\frac{\mpl\,(4\,\pi\,\rh^2)}{16\,\pi\,\lp}
\simeq
\lp\,\mpl\,\left(\frac{M}{\Meff}\right)^{\beta}
\ ,
\label{sS}
\ee
where
\be
\Meff=
\lp\left[\frac{1}{4}\left(\frac{\Mpl}{\md}\right)^{\alpha+\beta+2}\right]^{-\frac{1}{\beta}}
\ .
\ee
The area law then implies that the degeneracy of a black hole is
counted in units of $\Meff$ and
a black hole is classical if its mass is much larger than $\Meff$,
which implies that $\Meff$ must be no larger than $\md$ in order to have
TeV-scale black holes.
Since $\beta>0$, $\Meff\lesssim\md$ implies
\be
\alpha\gtrsim -2
\ ,
\label{alpha-2}
\ee
for all values of $\beta$.
For $\beta\not=1$, one then has that $\rc=L$ corresponds to a critical mass
\be
\Mc=
\left[L\,\lp^{~\beta-2}\left(\frac{\Mpl}{\md}\right)^{-\alpha-\beta-2}
\right]^{\frac{1}{\beta-1}}
\ .
\label{mc}
\ee
Further, for $\beta\not=2$, the condition that $\rc=\rh$ leads to
\be
M\simeq
\Mh
\equiv
\lp\left(\frac{\Mpl}{\md}\right)^{-\frac{\alpha+\beta+2}{\beta-2}}
\ ,
\label{mh}
\ee
whereas for $\beta=2$ is already assured by Eq.~\eqref{alpha-2}.
If we look at the conditions~\eqref{rcc} and \eqref{sbh}
for $\beta\neq 1$ and $\beta\neq 2$, so that $\mc$ and $\mh$ are
properly defined as above, we notice that the condition~\eqref{rcc}
implies
\be
M^{\beta-1}
\ll
L\,\lp^{~\beta-2}\,\left(\frac{\Mpl}{\md}\right)^{-\alpha-\beta-2}
\ ,
\label{rc_cond}
\ee
and we have the two cases
\be
&&
M\gtrsim\Mc
\ ,
\qquad
{\rm for}\ 0<\beta<1
\label{M>Mc}
\\
\nonumber
\\
&&
M\lesssim\Mc
\ ,
\qquad
{\rm for}\ \beta>1
\ .
\label{M<Mc}
\ee
Similarly, one can analyze the lower bound in Eq.~\eqref{sbh}.
Since we are only interested in small black holes, we
assume that the condition~\eqref{rc_cond} is satisfied.
Below the critical radius, where the tidal term dominates,
the horizon radius~\eqref{rhq} can be approximated by the
tidal component and $\rh\ll\rc$ yields
\be
M^{\beta-2}
\gg
\lp^{~\beta-2}\left(\frac{\Mpl}{\md}\right)^{-\alpha-\beta-2}
\ .
\ee
We again have two separate cases:
\be
&&
M\lesssim\Mh
\ ,
\qquad
{\rm for}\ 0<\beta<2
\label{betall2}
\\
\nonumber
\\
&&
M\gtrsim\Mh
\ ,
\qquad
{\rm for}\ \beta>2
\ .
\label{betagg2}
\ee
A detailed discussion on the constraints on the parameters $\alpha$
and $\beta$ for small black holes can be found in Ref.~\cite{cfhm}.
The overall conclusion is that, for any positive value of the parameter $\beta$,
the parameter $\alpha$ is constrained in the range
\be
-2\lesssim\alpha\lesssim-1.1
\ .
\label{fc}
\ee
\par
Finally, let us remark that the parameterization~\eqref{Q} for $Q$ in terms of
$\alpha$ and $\beta$ is not necessary for the present investigation, and
was only recalled here to establish a connection with previous
results~\cite{CH,cfhm,cfh}.
\subsection{Analytical and numerical results}
Starting from the brane metric~\eqref{tidal}, we ran our propagating algorithm to obtain
the bulk metric with terms up to $n\le 29$ for any $M$, $Q$ and $\sigma$.
These analytical expressions can then be used to study how the axis of cylindrical
symmetry and the horizon~\eqref{rhq} propagate into the bulk numerically.
In particular, the axis of cylindrical symmetry should start on the brane at $r=z=0$ and,
in the $(r,z)$-plane, be again represented by the line $R^2(r,z^{\rm axis}(r))=0$
to the right of which $R^2>0$, whereas the region to the left of $z=z^{\rm axis}(r)$,
having $R^2<0$, is unphysical.
Analogously, a proper horizon should start from the brane horizon at $r=\rh$, $z=0$
and be represented by a line $N(r,z^{\rm h}(r))=0$ inside the region to the right of
the axis $z=z^{\rm axis}(r)$.
As in Ref.~\cite{cm}, we are searching for those cases in which both the axis and
the horizon are regular and the horizon closes towards the axis along the extra
dimension.
We shall again call  the ``tip'' the point where axis and horizon cross.
Existence of the tip would thus signal the fact that the real singularity is enclosed
within the horizon and the corresponding brane metric is a viable
candidate for a brane-world black hole.
\par
For example, the metric elements to order $n=5$ are simple enough for displaying, namely
\be
N_5(r,z)
&\!=\!&
e^{-\sigma\, z}
\left[1-\frac{2\,M}{r}-\frac{Q}{r^2}
+\frac{\left(e^{-\sigma\, z}-1\right)^2\,Q}{\sigma^2\, r^4}
-\frac{2\,\left(e^{-\sigma\, z}-1\right)^2\,Q\,M}{\sigma^2\, r^5}
\right]
\label{N5}
\\
A_5(r,z)
&\!=\!&
e^{-\sigma\, z}
\left\{1
+\frac{2M}{r}-\frac{4\,M^2+Q}{r^2}+\frac{4\,M\left(2\,M^2+Q\right)}{r^3}
\right.
\nonumber
\\ 
& &\phantom{e^{-\sigma z}~~}
\left.
+\frac{\left(e^{-\sigma z}-1\right)^2\, Q
+\sigma^2\left(16\,M^4+12\,M^2\, Q+Q^2\right)}{\sigma^2\, r^4}
\right\}
\nonumber
\\
& &\phantom{e^{-\sigma z}~~}
\left.
+\frac{2\,M
\left[\left(e^{-\sigma\, z}-1\right)^2\, Q+\sigma^2\left(16\,M^4+16\,M^2 Q+3\,Q^2\right)\right]}
{\sigma^2\,r^5}
\right\}
\label{A5}
\\
R_5^2(r,z)
&\!=\!&
r^2\, e^{-\sigma\, z}\left[1-\frac{\left(e^{-\sigma\, z}-1\right)^2\, Q}{\sigma^2\, r^4}\right]
\ .
\label{R5}
\ee
From Eq.~\eqref{R5}, the axis of cylindrical symmetry at order $n=4$ is then given by
\be
z_4^{\rm axis}(r)
=
\frac{1}{\sigma}\,\ln\left(1+\frac{\sigma\, r^2}{\sqrt{Q}}\right)
\ ,
\label{zr5}
\ee
and, analogously, from Eq.~\eqref{N5}, the horizon at $n=4$ is located at
\be
z_{4}^{\rm h}(r)
=
\frac{1}{\sigma}\,\ln\left(1+\sigma\, r\,\sqrt{1+\frac{2\,M\,r}{Q}-\frac{r^2}{Q}}\right)
\ .
\label{zn5}
\ee
Finally, the horizon intersects the axis of symmetry at
\be
\begin{array}{l}
r^{\rm tip}_4
=
\strut\displaystyle\frac{1}{2}\left(M+\sqrt{M^2+2\,Q}\right)
\\
\\
z^{\rm tip}_4
=
\strut\displaystyle\frac{1}{\sigma}\,
\ln\left[1+\frac{\sigma}{4\,\sqrt{Q}}\left(M+\sqrt{M^2+2\,Q}\right)^2\right]
\ .
\end{array}
\ee
Note that $r^{\rm tip}_4$ and $r^{\rm tip}_4$ are always real and no constraint
is therefore derived for $M$, $Q$ and $\sigma$ at $n=4$ from requiring
the existence of a ``tip''.
However, looking at the term of order $n=5$ in Eq.~\eqref{N5},
one realizes that higher order terms will appear with possible alternating signs and
the picture become more involved, as we are now going to show.
\par
In fact, the propagating algorithm produces very cumbersome expressions rather quickly
as one raises the order $n$ of the multipole expansion.
In order to visualize the bulk structure, it is then convenient to consider
specific numerical values for the parameters $M$, $Q$ and $\sigma$ in the
range of interest for our study.
For this purpose, we also define the dimensionless quantities
$\bar{Q}=Q/\lp^2$, $\bar{M}=M/\lp$ and $\bar{\sigma}=\lp\,\sigma$.
The condition~\eqref{csigmanob} then reads
\be
\bar M\ll \bar\sigma^{-1}
\ .
\label{csigma}
\ee
Further, the relation~\eqref{Q} can be rewritten as
\be
\bar{Q}=\bar{M}^{\beta}\left(\frac{\Mpl}{\md}\right)^{\alpha+\beta+2}
\ .
\label{Qbar}
\ee
where $\beta>0$ and $\alpha$ satisfies Eq.~\eqref{fc}.
We therefore find that
\be
\bar{Q}\gg\bar{M}
\ ,
\label{QM}
\ee
and microscopic tidal-charged black holes should look very similar to five-dimensional
Schwarzschild black holes, in agreement with perturbative calculations~\cite{katz}.
\par
From the metric elements evaluated to order $n=29$,
we found that the horizon is well-behaved and the bulk is regular provided
\be
\bar M\lesssim \bar Q\ll\bar \sigma^{-1}
\ .
\label{tc}
\ee
whereas the bulk shows some pathological behaviors when one of the conditions
in Eq.~\eqref{tc} fails, namely for $\bar Q<\bar M$ or $\bar\sigma^{-1}<\bar Q$.
Note that Eq.~\eqref{tc} implies both~\eqref{csigma} and \eqref{QM},
but the latter is required {\em a priori\/}, in order to have small black holes
(not for regularity purposes).
\par
\begin{table}[t]
\begin{center}
\begin{tabular}{|c c||c|c|c|c|c|c|c|}
\hline
& $\bar Q$ & $10$ & $1$ & $10^{-1}$ & $10^{-2}$ & $10^{-3}$ & $10^{-4}$ & $10^{-5}$
\\
$\bar M$ & & & & & & & &
\\
\hline
\hline
$10^{-1}$ & & MsA MH & MsA OH & MH & MsA MH & MsA MH & MsA MH& MsA MH
\\
\hline
$10^{-2}$ & & MsA OH & MsA OH & MH & OH & MH  & MsA MH & MsA MH
\\
\hline
$10^{-3}$ & & MsA MH & MsA MH & R & R & MlA & OH &  MH
\\
\hline
$10^{-4}$ & & MsA OH & MsA MH & R & R & R & R & MlA OH
\\
\hline
$10^{-5}$ & & OH & MH & R & R & R & R  & MlA OH
\\
\hline
\end{tabular}
\end{center}
\caption{Bulk pathologies (or absence thereof).}
\label{table}
\end{table}
The above result was obtained by performing an extended survey in the space of
the parameters $\bar M$ and $\bar Q$ with $\bar\sigma^{-1}=1$.
A sample is shown in Table~\ref{table}, where the entries represent the peculiar
features of each case, which we describe in details below.
Before that, we remark that we were not able to establish a one-to-one
correspondence between $Q$ and $M$, but we rather obtained a range
of possible values for $Q$ given a value of $M$.
This is a consequence of the multipole expansion: the cases with regular bulk
(labelled as R in Table~\ref{table}) for a given value of $M$ represent different
brane metrics whose bulks do not show any pathology within the level of precision
allowed by our method.
In particular, we expect each value of $Q$ corresponds to a different shape of
the source, but when the sources are localized very near $r=z=0$, the multipole
expansion cannot discern them and we are therefore not able to rule out cases
for which, e.g., the source spreads off-brane and is not point-like.
\subsubsection{Regular bulk (R)}
\label{R}
\begin{figure}[t!]
\centering
\includegraphics[scale=0.7]{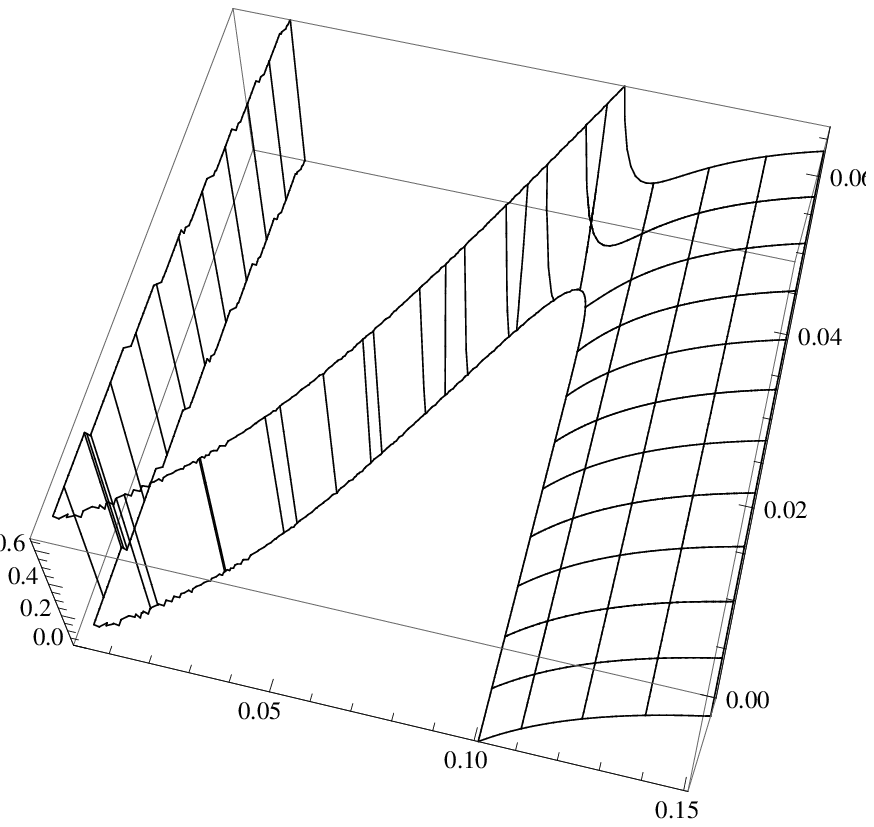}
$\quad$
\includegraphics[scale=0.7]{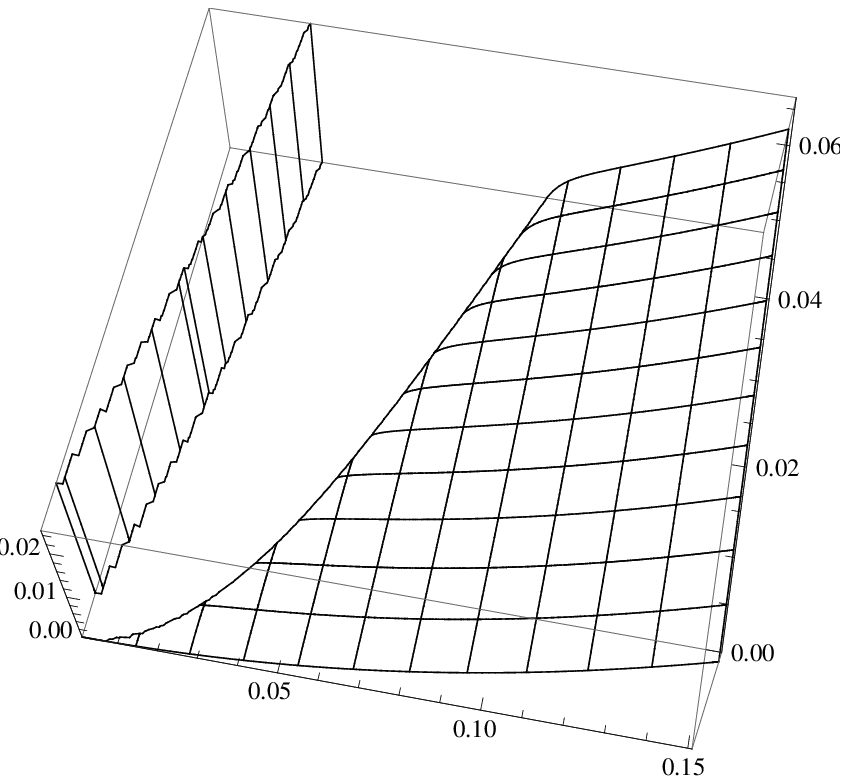}
\caption{Left panel: Plot of $N=-g_{tt}(r,z)\ge 0$.
Horizon is line $N=0$.
Right panel: Plot of $R^2(r,z)\ge 0$.
Axis of cylindrical symmetry is line $R^2=0$ right of which $R^2>0$
and space to left of axis ($R^2<0$) is unphysical.
(We set $\bar\sigma=1$, $\bar M=10^{-3}$ and $\bar Q=10^{-2}$)
}
\label{gttrz}
\end{figure}
\begin{figure}[h!]
\centering
\includegraphics[scale=0.7]{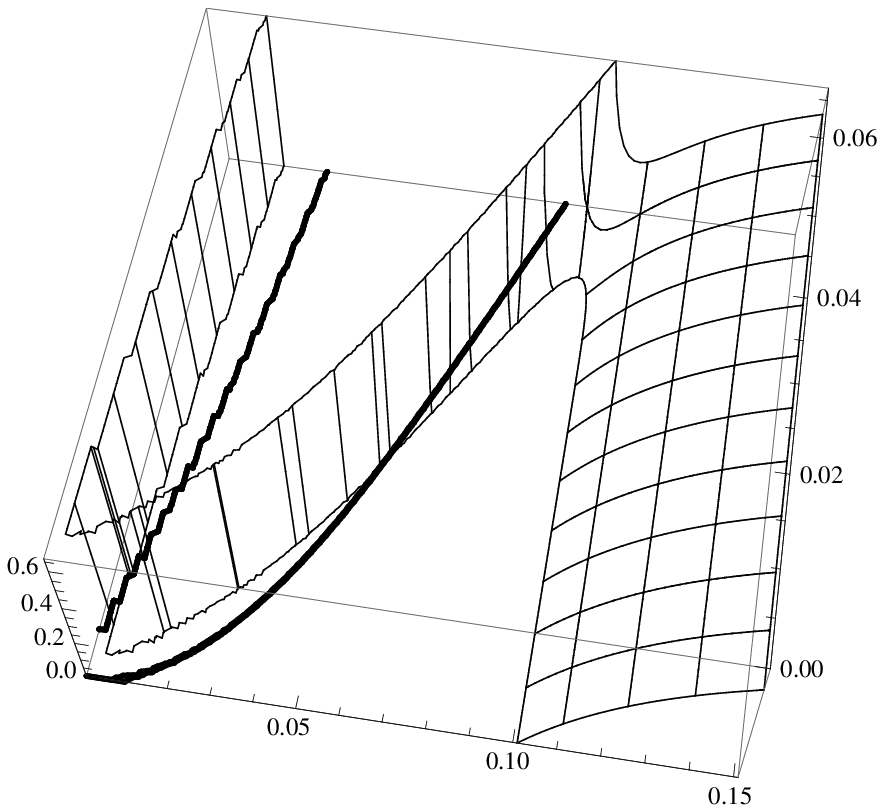}
$\quad$
\includegraphics[scale=0.7]{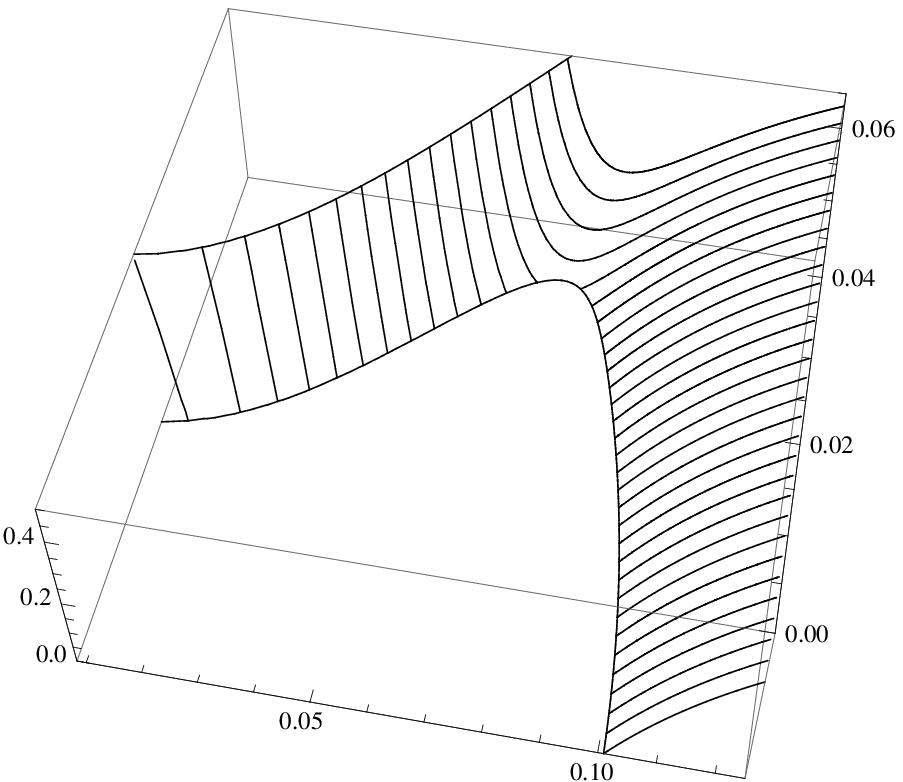}
\caption{Left panel:
Plot of $N=-g_{tt}(r,z)$ and $R^2(r,z)=0$ (thick line) from Fig.~\ref{gttrz}.
Right panel:
Plot of $N=-g_{tt}(R,z)\ge 0$.
Horizon is at $N=0$ and axis of cylindrical symmetry at $R=0$.
(We set $\bar\sigma=1$,
$\bar M=10^{-3}$ and $\bar Q=10^{-2}$)}
\label{gttRz}
\end{figure}
We denote with R the good candidates which show no pathology.
An example is given in Figs.~\ref{gttrz}-\ref{gttRz}  for $\bar\sigma=1$,
$\bar M=10^{-3}$ and $\bar Q=10^{-2}$.
From the left panel in Fig.~\ref{gttrz}, we see a horizon at $r\simeq 0.1$
on the brane which closes towards smaller values of $r$ in the bulk.
The right panel shows $R^2$ plotted in the $(r,z)$-plane as well.
The space to the left of the line where $R^2=0$ is unphysical.
As mentioned before, for the horizon to close in the bulk, it must cross the axis
of cylindrical symmetry at a tip point, where $R^2=0$.
The left panel in Fig.~\ref{gttRz} shows that there is indeed such a point.
Finally, the right panel in Fig.~\ref{gttRz} displays $N$ directly as a function
of $(R,z)$ and shows that the horizon approaches the axis where $R=0$.
\par
A final remark is in order.
Some of the plots for regular cases show an axis that does not
exactly start from $r=z=0$.
This can be due to the limitations of the multipole expansion
and necessarily limited numerical precision.
We therefore do not consider this feature as evidence of
any pathologies.
\subsubsection{Multiple axis (MsA and MlA)}
\label{MA}
\begin{figure}[t!]
\centering
\includegraphics[scale=0.7]{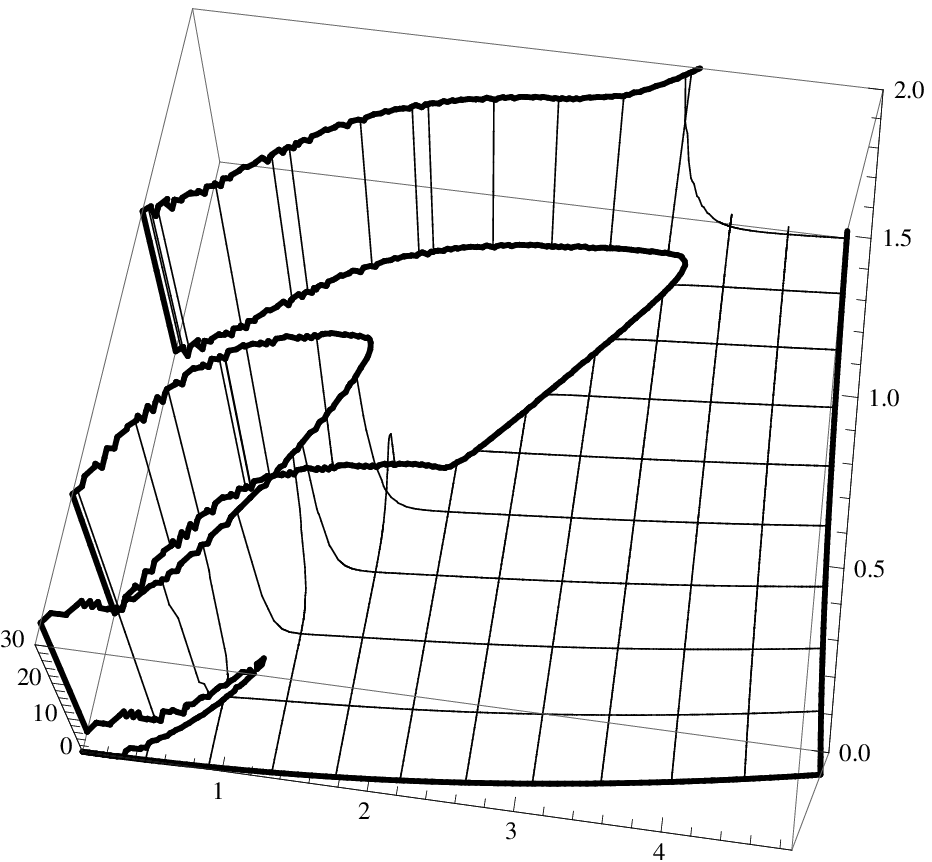}
$\qquad$
\includegraphics[scale=0.7]{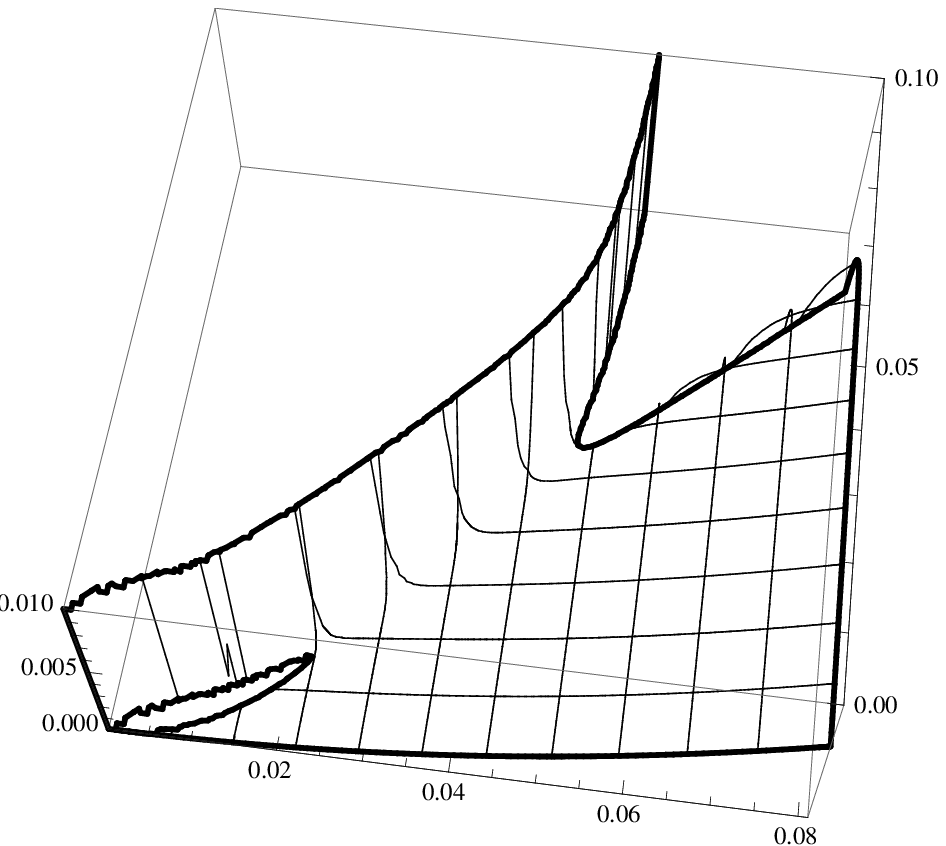}
\\
MsA
\hspace{7cm}
MlA
\caption{Left panel: Plot of $R^2(r,z)\ge 0$
for $\bar M=10^{-1}$ and $\bar Q=10$.
One axis is in bottom left corner and one above it.
Right panel: Plot of $R^2(r,z)\ge 0$
for $\bar M=\bar Q=10^{-3}$.
One axis is in the left bottom corner and one
appears in the top right.
(We set $\bar\sigma=1$.)
}
\label{ma}
\end{figure}
There are cases which show more, apparently disconnected,
lines $z=z^{\rm axis}(r)$ along which $R^2=0$, as shown
by the two examples given in Fig.~\ref{ma}.
In the left panel, we see a first line $z=z^{\rm axis}(r)$ close to
the left bottom corner, and what looks like a larger copy of it still
emanating from $r=0$.
In this case, it is difficult to assess whether the two lines are
disconnected or rather meet around $r=0$, since the multipole expansion
is not reliable for small $r$.
We denote this first case as Multiple short Axis (MsA).
In the right panel, we see a similar $z=z^{\rm axis}(r)$
near the bottom left corner which winds around $r=z=0$,
and a second, V-shaped $z=z^{\rm axis}(r)$ in the right
top corner.
These two lines appear clearly disconnected and the larger one
is well separated from the brane axis at $r=z=0$.
We denote this case as Multiple large Axis (MlA).
\par
In general, we consider the presence of a multiple axis of the first kind
(MsA) as a warning sign that our approach might not be able to describe
the space-time reliably around the axis.
However, a multiple axis of the second kind (MlA) makes the bulk irregular
and unacceptable.
\subsubsection{Multiple horizon (MH)}
\label{MH}
With Multiple Horizon (MH) we denote cases in which there are more lines
$z=z^{\rm h}(r)$ in the region where $R^2>0$.
We consider the appearance of any $z=z^{\rm h}(r)$, beside the one generating
from $r=\rh$, as a sufficient reason to discard the case.
Ideally, such situations should contain extra sources beside the point-like tidal black hole.
An example is given in the left panel of Fig.~\ref{mh}.
Near the left bottom corner, we see the brane horizon which closes towards the
axis as in the regular cases.
However, a second horizon appears in the bulk, near the right top corner,
where $R^2>0$.
\subsubsection{Open horizon (OH)}
\label{OH}
When the horizon starting from $r=\rh$ at $z=0$ does not close
towards the axis, we have an Open Horizon (OH).
This is again considered a sufficient reason to discard the case, since it
presumably signal the source at $r=z=0$ is not point-like but extends off-brane,
or other off-brane sources are anyway present.
An example of this behavior is shown in the right panel of Fig.~\ref{mh}.
\begin{figure}[t!]
\centering
\includegraphics[scale=0.7]{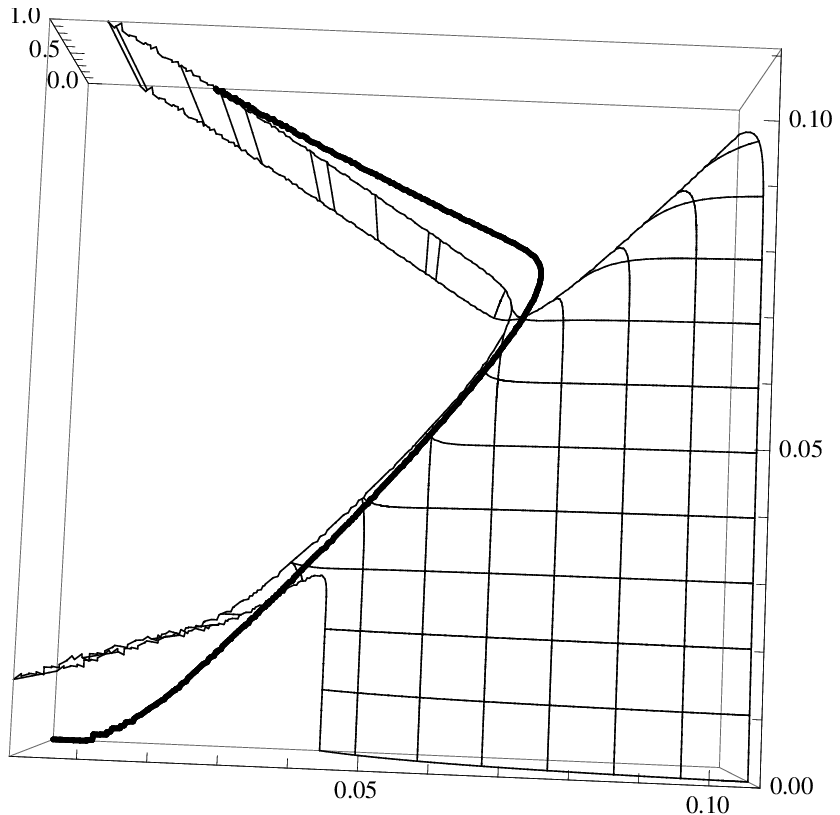}
$\qquad$
\includegraphics[scale=0.7]{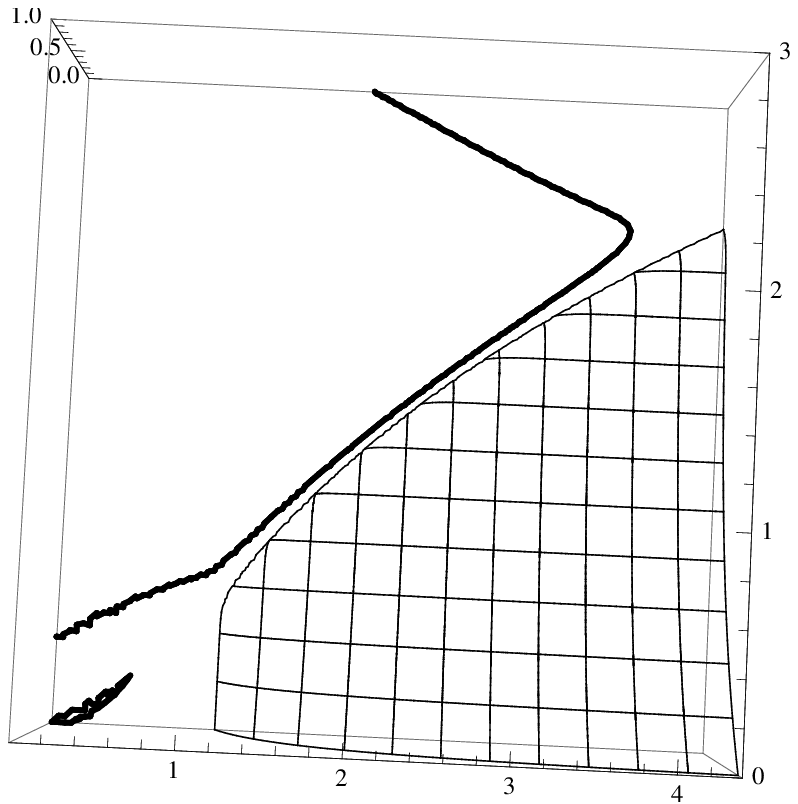}
\\
MH
\hspace{6cm}
OH
\caption{Left panel: Plot of $N=-g_{tt}(r,z)\ge 0$ and $R^2(r,z)= 0$ (thick line)
for $\bar M=10^{-2}$ and $\bar Q=10^{-3}$.
Horizon generating from brane is near bottom left corner and
crosses axis.
Second horizon appears near right top corner starting from axis.
Right panel: Plot of $N=-g_{tt}(r,z)\ge 0$ and $R^2(r,z)=0$ (thick line)
for $\bar M=10^{-1}$ and $\bar Q=1$.
Horizon starts from brane and never crosses axis.
(We set $\bar\sigma=1$)
}
\label{mh}
\end{figure}
\section{Conclusions}
\setcounter{equation}{0}
In this paper, we studied the bulk corresponding to tidal charged brane-world
black holes for the case of microscopic masses within the energy range
of the LHC.
Such black holes are characterized by the four-dimensional ADM mass $\bar M$
and tidal charge $\bar Q$~\cite{maartens}.
To these (dimensionless) parameters, one must also add the brane density
$\bar\sigma$, and the conditions for black holes to be in the TeV-range are
then given in Eqs.~\eqref{csigma} and~\eqref{QM}.
\par
In order to reconstruct the bulk, we employed a propagating algorithm which
makes use of the three-dimensional multipole expansion of any brane metric
of choice, and then allows one to determine analytically the corresponding
five-dimensional metric elements.
Given a generic brane-world metric, it is possible that singularities of various
kind appear off the brane, thus signaling that the starting four-dimensional metric
is not a good candidate in RS.
For the specific family of brane-world metrics of interest here, we found that
the bulk structure is regular for $\bar M$, $\bar Q$ and $\bar \sigma^{-1}$
satisfying Eq.~\eqref{tc}.
By regular we mean that the brane-world horizon is propagated into the bulk
smoothly and closes towards the axis of cylindrical symmetry along the extra
dimension for $\bar M\lesssim \bar Q\ll\bar \sigma^{-1}$ [see Section~\ref{R}].
On the contrary, when any of these conditions~\eqref{tc} is violated
[keeping Eq.~\eqref{QM} in order for the black hole to be small],
the bulk shows very peculiar causal structures described
in Sections~\ref{MA}-\ref{OH}.
\par
By comparing this result with the form of the tidal charge~\eqref{Q} first given
in Ref.~\cite{cfhm}, we found that the values of the parameters $\alpha$ and $\beta$
which ensure~\eqref{tc} are indeed the same that were previously selected
in Ref.~\cite{cfhm} on purely phenomenological grounds.
We wish to stress, nonetheless, that the results presented here do no depend on
the specific form chosen to relate $Q$ to $M$ and other parameters in the
model, since they were all kept free when generating the bulk metric.
\section*{Acknowledgments}
We thank B.~Harms for stimulating discussions.
R.C.~is supported by INFN grant BO11.
\end{document}